\documentclass[reprint,aps,prx,showkeys]{revtex4-2}

\usepackage{amsmath,amssymb}
\usepackage{color}
\usepackage{graphicx}
\usepackage{dcolumn}
\usepackage[normalem]{ulem}
\usepackage[shortlabels]{enumitem}

\usepackage[colorinlistoftodos]{todonotes}

\usepackage{fancyhdr}

\begin{document}

\preprint{APS/123-QED}

\title{Transverse Momentum
Exchange in Optical Vector Beam Propagation: \\Theory and Experiment}


\author{J. M. Nichols and D. V. Nickel}
\affiliation{
U.S. Naval Research Laboratory, 4555 Overlook Ave. SW., Washington D.C. 20375\\
}

\author{F. Bucholtz} 
\affiliation{Jacobs Technology, Inc., 2551 Dulles View Drive Herndon, VA 20171\\
}

\date{\today}

\begin{abstract}
In this work we present a model for the paraxial propagation of vector beams. Of particular importance is the appearance of a new transverse momentum term proportional to the linear polarization angle gradient imparted to the beam during the beam preparation process. The model predicts that during propagation, this new momentum will be exchanged for classical transverse momentum, thereby causing the beam to accelerate. We observe the momentum exchange in experiment by designing the polarization profile in such a way as to cause the beam centroid to follow a parabolic path through free space.
\end{abstract}


\maketitle

\section{Introduction}
One of the more fundamental set of predictions that can be made in optical physics is the magnitude, phase, and polarization of a beam at a location downrange from an aperture.  Such predictions form the foundation of technology development for imaging, spectroscopy, countermeasures, communications, ranging, etc.  Despite the hundreds of years devoted to the associated models, the field continues to see the emergence of new physical phenomena that in some cases have challenged these models and suggested modifications.  Examples include beams with novel phase \cite{Milione:12}, amplitude \cite{Zhao:22}, and polarization \cite{Wang:21,Nichols:22,Nichols:23} distributions. 
These novel ``structured light fields'' are changing our collective understanding of fundamental quantities such as optical momentum \cite{Berry:09,Bekshaev:11}, the Poynting vector \cite{Paganin:98,Nugent:00} and the Maxwell stress tensor \cite{Nieto:22}.  
\pagestyle{plain} 

In this work we develop the physics underlying the paraxial propagation of beams with spatially inhomogeneous distributions of linear polarization.  Importantly, the transverse momentum vector of such beams is shown to include a new term proportional to the beam's polarization angle gradient. We describe how this new polarization momentum is effectively stored and then subsequently expended on propagation to produce transverse beam acceleration - a necessary consequence of momentum conservation.  We further identify and highlight the critical role diffraction plays as a dissipative mechanism for this stored momentum, thereby driving the momentum exchange.  Next, we predict specific, curved beam paths corresponding to particular polarization profiles and then validate those predictions in experiment.  Indeed, such a beam is seen to follow precisely the predicted parabolic path, yielding centroid deflections of several millimeters  at 35m propagation distance for a 1.5 $\mu$m wavelength monochromatic beam.

We also provide an alternative derivation of the model based on the principle of stationary action.  By viewing polarization angle and dynamical phase as equivalent contributions to the Lagrangian density, we demonstrate that polarization-gradient beams constitute a symmetry in the context of field theory.  The term that offers control over the beam path is then immediately recognized as a ``symmetry-breaking'' term. Collectively, these results comprise a fundamentally new understanding of paraxial beam propagation that predicts novel, previously unobserved behavior, and includes standard paraxial beam theory as a subset.

The model begins with the vector paraxial wave equation 
\begin{align}
\left[
-i2k_0 \partial_z +\nabla^2_X
\right]
\left\{E_X(\vec{x},z),E_Y(\vec{x},z)\right\}
=0
\label{eqn:paraWave}
\end{align}
which governs the complex amplitude of the transverse, monochromatic electric field $\vec{E}(\vec{x},z,t)=\{E_X(\vec{x},z),~E_Y(\vec{x},z)\}\exp[i(\omega t-k_0z)]$. The notation $\nabla_X(\cdot)\equiv \{\partial_x(\cdot),\partial_y(\cdot)\}$ denotes the transverse gradient operator.     

Equation (\ref{eqn:paraWave}) can be constructed directly from Maxwell's equations assuming 1) propagation occurs in free space 2) propagation proceeds mainly in the preferred direction $z$ (that is, no backscatter and deviations from the $z-$axis of less than $\approx 0.1$ rad) with phase accumulating in that direction at a rate governed by wavenumber $k_0$, and 3) over $z-$distances of a wavelength or so, variations in the field amplitude are much smaller than the amplitude itself (the slowly-varying envelope approximation).  
Solutions to Eq.(\ref{eqn:paraWave}) therefore comprise the spatially-dependent, slowly-varying field amplitudes $E_X(\vec{x},z),E_Y(\vec{x},z)$ associated with temporal frequency $\omega=ck_0$.  Note that in free space, Maxwell's equations mandate the divergence of the electric field be zero, $\nabla\cdot\{E_X(\vec{x},z),E_Y(\vec{x},z),E_Z(\vec{x},z)\}=0$. Enforcing this constraint under assumption 3) then requires that the $\hat{z}$ component of the electric field be given by $E_Z(\vec{x},z)=i/k_0\nabla_X\cdot\{E_X(\vec{x},z),E_Y(\vec{x},z)\}$ \cite{Lax:75,Berry:09}. The dynamics of interest (hence our model) is therefore governed entirely by the spatial variation of the transverse field amplitudes.  Moreover, for the optical beams studied in this work the power in the $\hat{z}$ field component is significantly less than the power in the transverse components \cite{Lax:75}.

\section{Free-space propagation of vector beams\label{sec:freespace}}

Consistent with the stated assumptions, we presume the commonly-used transverse vector electric field model red{\cite{Chen:18,Zhao:22b}}
\begin{align}
&\{E_X(\vec{x},z),E_Y(\vec{x},z)\}=\nonumber \\
&\qquad\qquad\rho^{1/2}(\vec{x},z) e^{-i\phi(\vec{x},z)}\left\{
\cos{\gamma(\vec{x},z)},\sin{\gamma(\vec{x},z)}\right\}
\label{eqn:efield}
\end{align}
and substitute into (\ref{eqn:paraWave}).  This transforms the complex Eq.(\ref{eqn:paraWave}) into the real-valued system of equations (see Appendix \ref{sec:derivation})
\begin{subequations}
    \begin{align}    \partial_z\rho(\vec{x},z)+\nabla_X\cdot\left[\rho(\vec{x},z)\vec{v}(\vec{x},z)+\rho(\vec{x},z)\vec{\Omega}(\vec{x},z)\right]&=0
    \label{eqn:continuity}
    \end{align}
    \begin{align}
    \frac{D \vec{v}(\vec{x},z)}{Dz}&=-\left(\vec{\Omega}(\vec{x},z)\cdot\nabla_X\right)\vec{\Omega}(\vec{x},z)\nonumber \\
    &\qquad\qquad+\frac{1}{2k_0^2}\nabla_X\left(\frac{\nabla_X^2\rho^{1/2}(\vec{x},z)}{\rho^{1/2}(\vec{x},z)}\right)  
    \label{eqn:velocity}
    \end{align}
    \begin{align}
    \frac{D\vec{\Omega}(\vec{x},z)}{Dz}+\left(\vec{\Omega}(\vec{x},z)\cdot\nabla_X\right)\vec{v}(\vec{x},z)=0
    \label{eqn:vorticity}
    \end{align}
    \label{eqn:model}
    \end{subequations}
where 
\begin{subequations}
\begin{align}\label{eqn:vDef}
    \vec{v}(\vec{x},z)=k_0^{-1}\nabla_X\phi(\vec{x},z) \\
    \vec{\Omega}(\vec{x},z)\equiv k_0^{-1}\nabla_X\gamma(\vec{x},z)
\end{align}
\end{subequations}
captures the spatially localized rate of change in the beam path in the transverse direction per unit change in the direction of propagation (i.e., a transverse ``velocity'') and the polarization angle gradient, respectively \cite{Nichols:22, Nichols:23}. The notation $D(\cdot)/Dz$ denotes total derivative \footnote{Recall the total derivative of function $f$ with respect to variable $s$ accounts for both the intrinsic partial derivative $\partial f/\partial s$ and also the transport path of the quantity $f$ through space. $Df/Ds=\partial f/\partial s+(\vec{v}\cdot\nabla) f$}. 

We first note that for homogeneously-polarized light, $\vec{\Omega}(\vec{x},z)=0$ and the model (\ref{eqn:model}) predicts the same amplitude and phase as scalar diffraction theory in the Fresnel approximation \cite{Nichols:19}.  In that case, one can exactly solve (\ref{eqn:paraWave}) independently for the two orthogonal field components $E_X(\vec{x},z)$ and $E_Y(\vec{x},z)$ and then simply superimpose (add) the results \cite{Goldsmith:98}. 
 However, it is well known that this approach does not, in general, solve the full vector wave equation due to coupling among vector components (see e.g. Goodman \cite{Goodman:68}, page 32 for a general statement or  Hall \cite{Hall:96} for a more specific example). Indeed, the model (\ref{eqn:model}) illustrates the tight coupling among the transverse vector components of phase and polarization gradients and, hence, the inability of scalar diffraction theory to capture the dynamics of inhomogeneously polarized beams in general. 
 
Equation (\ref{eqn:continuity}) is the familiar ``transport of intensity equation'' \cite{Paganin:98}, but it is augmented by a new transverse linear momentum component $\rho(\vec{x},z)\vec{\Omega}(\vec{x},z)$, namely, the product of the beam intensity and the scaled polarization gradient.

 Expression (\ref{eqn:velocity}) governs the optical path as determined by the transverse phase gradient.  The second term on the right hand side governs diffraction \cite{Nichols:19} while, importantly, the beam's trajectory is clearly also influenced by the polarization angle gradient.  In fact, through proper choice of $\vec{\Omega}(\vec{x},z)$ the beam can be caused to accelerate and follow a parabolic path through free-space.  This method of ``beam bending'' was demonstrated initially in \cite{Nichols:22} and will be explored further in section \ref{sec:experiment}.

Interestingly, equation (\ref{eqn:vorticity}) is well known in fluid mechanics as the vorticity transport equation for inviscid flow with $\vec{\Omega}(\vec{x},z)$ playing the role of ``vorticity'' \cite{Saeed:22}.  This term is necessary for the conservation of momentum and, in fluid mechanics, (\ref{eqn:vorticity}) captures ``vortex stretching'', a phenomenon whereby velocity gradients change the rate of rotation of particles in a fluid flow.  In the optical case, $\vec{\Omega}(\vec{x},z)$ captures the variation of the optical axis in the transverse plane.  As the beam diffracts, the polarization gradient necessarily decreases as the transverse velocity is ``stretched'' (Eq. \ref{eqn:vorticity}), thereby decreasing the beam acceleration via (\ref{eqn:velocity}).  In this sense, diffraction dissipates the polarization gradient during propagation. 

As we will show next, the model (\ref{eqn:model}) describes an exchange  between the momentum stored within the polarization angle gradient and the transverse linear momentum as the beam propagates. The former is generated during the beam preparation process and is then expended on propagation, due to diffraction, which causes the beam to accelerate via a corresponding increase in the classical transverse linear momentum.  Once the stored momentum is expended, the model predicts the beam will cease to ``bend'' and will simply follow a straight path governed by the standard diffraction theory, albeit at some non-zero angle relative to the $z$ direction. 

\section{Momentum Conservation and Momentum Exchange \label{sec:momentum}} 

The model (\ref{eqn:model}) can alternatively be written in the ``conservation form'' commonly used in continuum mechanics.  
Following the basic steps outlined in Appendix \ref{sec:derivation},
we can express (\ref{eqn:model}) as
\begin{subequations}
\begin{align}   \partial_z\rho(\vec{x},z)+\nabla_X\cdot\left[\vec{P}(\vec{x},z)\right]=0,
\label{eqn:continuityEqn}
\end{align}
\begin{align}
    \partial_z \vec{P}(\vec{x},z)+\nabla_X\cdot\left[\vec{P}(\vec{x},z)\otimes \frac{\vec{P}(\vec{x},z)}{\rho(\vec{x},z)}\right]=-\nabla_X\cdot {\bf P}(\vec{x},z)
    \label{eqn:momentumEqn}
\end{align}
\label{eqn:conserve}
\end{subequations}
where the $\otimes$ symbol denotes vector outer product and $\vec{P}(\vec{x},z)$ is a generalization of the transverse optical Poynting vector 
\begin{align}
\vec{P}(\vec{x},z)=&\rho(\vec{x},z)\Bigl[\vec{v}(\vec{x},z)+\vec{\Omega}(\vec{x},z)\Bigr].
\label{eqn:Poynting}
\end{align}
The tensor ${\bf P}(\vec{x},z)$ plays the role of a diffractive ``pressure'' as will be discussed shortly.  This form of the model combines the phase and polarization vector fields into a single quantity, highlighting the fact that, if only intensity is measured downrange, the contribution from these two fields cannot be distinguished. Equations (\ref{eqn:continuityEqn}) and (\ref{eqn:momentumEqn}) express (spatially) local conservation of intensity and transverse linear momentum, respectively.  Integrating these expressions over the transverse plane (see Appendix \ref{sec:conserve}) 
gives the corresponding global conservation statements.  Note the momentum density in the direction of propagation is simply $\rho(\vec{x},z)$ for this paraxial model \cite{Bekshaev:07} and does not play a role in the transverse dynamics. 

We can also define the transverse Maxwell stress tensor in the paraxial limit,
\begin{align}
    {\boldsymbol \sigma}(\vec{x},z)={\bf P}(\vec{x},z)+\vec{P}(\vec{x},z)\otimes\frac{\vec{P}(\vec{x},z)}{\rho(\vec{x},z)}
\end{align}
in which case (\ref{eqn:momentumEqn}) further condenses to
\begin{align}
    \partial_z\vec{P}(\vec{x},z)+\nabla_X\cdot{\boldsymbol\sigma}(\vec{x},z)=0.
    \label{eqn:Lorenz}
\end{align}
The expression (\ref{eqn:Lorenz}) mirrors that which is often derived from the familiar Lorenz force law \cite{Davis:20}.   

This form of the model, and the structure of Eq. (\ref{eqn:Poynting}) clearly suggests two types of local, transverse optical momentum: the conventional $\rho(\vec{x},z)\vec{v}(\vec{x},z)$ (see e.g., \cite{Bekshaev:07b}, Eqn. 20) and a second type, $\rho(\vec{x},z)\vec{\Omega}(\vec{x},z)$.  Both are required to write the conservation form of our system (\ref{eqn:momentumEqn}) and thus demonstrate that momentum is conserved.  Moreover, just as the classical momentum is proportional to a phase gradient, so too is this new type of momentum.   We show in Appendix \ref{sec:vecPhase} that the initial transverse momentum $\rho(\vec{x},0)\vec{\Omega}(\vec{x},0)$ is directly related to a geometric, Pacharatnam-Berry (PB) phase gradient \cite{Pancharatnam:56,Berry:84,Bliokh:19} that accompanies the series of polarization transformations in the beam preparation process  (see section \ref{sec:experiment} or ref. \cite{Nichols:22}). It is therefore not surprising that $\rho(\vec{x},z)\vec{\Omega}(\vec{x},z)$ captures a type of transverse optical momentum.  

As we will show in experiment, we have $\rho(\vec{x},0)\vec{v}(\vec{x},0)=0$ (collimated light at the exit aperture) and $|\rho(\vec{x},0)\vec{\Omega}(\vec{x},0)|>0$. During propagation, diffraction will reduce the polarization gradient through (\ref{eqn:vorticity}) which will reduce the associated momentum.  Momentum conservation then mandates that the beam accelerate ($\rho(\vec{x},z)\vec{v}(\vec{x},z)$ increases).  This process continues until all of the initial polarization gradient momentum has been converted to $\rho(\vec{x},z)\vec{v}(\vec{x},z)$. 

Diffraction therefore plays a critical role in the dynamics.  For example, smaller beam diameters will lead to stronger diffraction which is predicted to dissipate the bending effect over shorter propagation distances. The dynamics we observe are therefore appropriately viewed as a {\it momentum exchange} whereby the initial momentum associated with the polarization gradient is converted to conventional transverse momentum thereby accelerating the beam.  

 To conclude this section, we note that an interesting byproduct of the formulation (\ref{eqn:conserve}) is the implicit definition of a diffractive force as the divergence of a pressure tensor \footnote{The outer product notation $\nabla_X\otimes\nabla_X f$ denotes $\begin{bmatrix}\partial_{xx}f & \partial_{xy}f \\ \partial_{yx}f & \partial_{yy}f\end{bmatrix}$}
\begin{align}
    {\bf P}&=-\frac{\rho(\vec{x},z)}{4k_0^2}\nabla_X\left(\frac{\nabla_X\rho(\vec{x},z)}{\rho(\vec{x},z)}\right)\nonumber \\
    &=-\frac{\rho(\vec{x},z)}{4k_0^2}\left(\vphantom{\vec{\Omega}}\nabla_X\otimes\nabla_X\log\left[\rho(\vec{x},z)\right]\right).
    \label{eqn:pressure}
\end{align}
Thus we arrive at an alternative view of diffraction, namely, that it stems from an internal ``pressure'' associated with the distribution of intensity. The pressure will clearly vanish for a uniformly distributed intensity, that is, a plane wave will not diffract.  However, by (\ref{eqn:momentumEqn}), we see that the transverse acceleration is driven by the negative of the divergence of this term, in other words, 
{\it diffraction drives uneven or "peaked`` intensity distributions toward flatter, more uniform distributions}, and the rate at which the flattening occurs depends on the logarithm of the distribution.  This is, to our knowledge, a new interpretation of diffraction.  We note that the ``Airy beam bending'' path found in the literature (see e.g., \cite{Latychevskaia:16}) can be modeled solely with this diffractive term, as we showed in \cite{Nichols:22}. The bending effect described in this work, however, is driven entirely by the polarization gradient $\vec{\Omega}(\vec{x},z)$ and therefore represents an entirely different  bending mechanism. 

\section{Alternative views of the model}

One can also arrive at the model (\ref{eqn:model}) through consideration of the Lagrangian density and the principle of stationary action.  Letting $\alpha(\vec{x},z)=\phi(\vec{x},z)+\gamma(\vec{x},z)$, the Lagrangian energy density can be written
\begin{align}
    \mathcal{L}&=-k_0\rho(\vec{x},z)\partial_z\alpha(\vec{x},z)-\frac{1}{2}\rho(\vec{x},z)|\nabla_X\alpha(\vec{x},z)|^2\nonumber \\
    &\qquad\qquad
    -\frac{|\nabla_X\rho(\vec{x},z)|^2}{8\rho(\vec{x},z)}.
    \label{eqn:Lagrangian}
\end{align}
Using the Euler-Lagrange equations to set the variation in $\alpha$ equal to zero yields directly Eq. (\ref{eqn:continuity}) while the variation in $\rho$ yields (upon taking the transverse gradient) the sum of Eq. (\ref{eqn:velocity}) and Eq. (\ref{eqn:vorticity}).  In the absence of a transverse gradient in $\gamma$, one recovers the hydrodynamic form of standard paraxial beam propagation \cite{Nichols:19}. 
Thus, going from a scalar (standard) to vector beam model simply involves the transformation $\phi\rightarrow \phi+\gamma$ in the Lagrangian density.  Moreover, this transformation is a {\it symmetry} in that, to leading order in $\gamma$, it adds  $D\gamma/Dz$ to the Lagrangian density which, as a consequence of Noether's Theorem, is zero. This mathematical statement (upon taking the transverse gradient) then matches Eq.  (\ref{eqn:vorticity}). Importantly, however, the transformation also results in the addition of the higher order term $\rho|\nabla_X\gamma|^2/2$ which breaks the symmetry and ultimately becomes the ``bending'' term in Eq. (\ref{eqn:velocity}).

That a polarization gradient and a phase gradient are on equal footing in the definition (\ref{eqn:Poynting}) and in the energy density (\ref{eqn:Lagrangian}) should not be entirely surprising. For example, if one represents the electric field (\ref{eqn:efield}) in the circular basis, $\gamma$ appears in the argument of a complex exponential (in the same manner as $\phi$). 

Additionally, the connection between polarization, optical momentum, and the aforementioned geometric PB phase has been known for some time (see e.g., \cite{Tiwari:92,Bomzon:01,Milione:12,Capasso:22,McDonnell:22,Jisha:21,Bliokh:15,Bliokh:19,Tiwari:24}).  The geometric PB phase is a fundamentally different quantity than the {\it dynamical} phase, $\phi(\vec{x},z)$ in that it is non-integrable \cite{Berry:84} and requires a vector (as opposed to scalar) electric field model.  This is yet another reason why scalar diffraction theory cannot capture the effect reported here. We have already discussed in section \ref{sec:momentum} how a spatially-dependent PB phase is accumulated during creation of the polarization gradient. 
The exact relationship between the PB phase gradient and the polarization angle gradient is provided in Appendix \ref{sec:vecPhase} 
for our specific experimental implementation.  Thus, our model can be viewed in terms of two distinct types of phase: dynamical and geometric.  

The new polarization gradient momentum possesses at least some of the properties of so-called ``hidden'' momentum (see e.g., \cite{Babson:09,Griffiths:12,Jimenez:22}).  For example, if one observed a transverse displacement of the beam centroid, but was unaware of the polarization gradient contribution, they would conclude there must be hidden source of momentum in the system. As noted in \cite{Jimenez:22}, ``the usual momentum density proportional to Poynting's vector is equal to the hidden momentum and cancels it'' which is precisely what is implied by Eq. (\ref{eqn:Poynting}) where $\rho(\vec{x},z)\vec{\Omega}(\vec{x},z)$ acts to balance the traditional transverse Poynting vector $\rho(\vec{x},z)\vec{v}(\vec{x},z)$.  

Based on our model structure, the term $\rho(\vec{x},z)\vec{\Omega}(\vec{x},z)$ is perhaps appropriately viewed as a {\it vector potential}.  In the language and notation of \cite{Konopinski:78,Martins:08} this term is referred to as a ``potential momentum'', ``available for exchange with kinetic momenta'' just as our model predicts.  Alternatively, this term plays the role of the density-dependent gauge potential in quantum mechanics \cite{Buggy:20}, as suggested by our Eq. (\ref{eqn:Lagrangian}). To be more specific, instead of (\ref{eqn:efield}), write the complex {\it scalar} potential
\begin{align}
    E(\vec{x},z)&=\rho^{1/2}(\vec{x},z)e^{-i\phi(\vec{x},z)}e^{-i\gamma(\vec{x},z)}
    \label{eqn:gauge}
\end{align}
which simply augments the traditional (scalar) model by the gauge function $\gamma(\vec{x},z)$ (the polarization angle in this case).  Substituting (\ref{eqn:gauge}) into the {\it scalar} paraxial wave equation again yields our model (\ref{eqn:model}).  The author of \cite{Buggy:20} goes on to note ``the action of a gauge potential can be mimicked by imparting a geometric phase onto the wavefunction'', precisely what we have done here in creating our polarization/Berry phase gradient (see again Appendix \ref{sec:vecPhase}). 
Thus, viewing a polarization angle as a geometric phase in the scalar beam model is equivalent to our initial treatment in the vector model (\ref{eqn:efield}).  The physical meaning of vector potentials has been debated for over 150 years \cite{Jackson:01}, and the model offered here offers a concrete example in the field of optics where the presence of such a potential predicts an effect -- bending -- that has been observed in experiment.   

Lastly, although the definition (\ref{eqn:Poynting}) occurs quite naturally in the transport model, it is not at all an obvious choice.  This difficulty was acknowledged by Berry \cite{Berry:09} who suggested multiple definitions of the transverse Poynting vector for paraxial vector light.  Bekshaev \& Soskin present a general definition of the Poynting vector for vector fields which includes both the standard (scalar) component $\rho(\vec{x},z)\vec{v}(\vec{x},z)$ and a ``Spin Flow Density'' (SFD) equal to the gradient of the third Stokes parameter \cite{Bekshaev:07b}.  For our linearly polarized beam the SFD is zero, although in an earlier work \cite{Nichols:23} we showed 
\begin{align}
  \rho(\vec{x},z)\vec{\Omega}(\vec{x},z)&=\lim_{\vec{x
'}\rightarrow\{0,0\}}\Big(-ik_0^{-1}\nabla_{X'}s_3(\vec{x},\vec{x'})\Big),
\label{eqn:stokes3}
\end{align}
that is, a spatial gradient in linear polarization is related to the transverse spatial gradient of the {\it generalized}  
Stokes parameter $s_3(\vec{x},\vec{x}')\equiv i\left[E_X(\vec{x}-\vec{x}'/2)E_Y^*(\vec{x}+\vec{x}'/2)-E_X^*(\vec{x}+\vec{x}'/2)E_Y(\vec{x}-\vec{x}'/2)\right]$ (easily verified by substituting Eq. \ref{eqn:efield} into Eq. \ref{eqn:stokes3}).  Alternatively, (\ref{eqn:stokes3}) can be written as an integral over transverse spatial frequencies, suggesting that changes in $\vec{\Omega}(\vec{x},z)$ yields changes in spatial frequencies of the beam, also the conclusion of \cite{Capasso:22}.  
Our observed effect occurs in free-space and is unrelated to the  optical Magnus \cite{Bliokh:04}, ``spin Hall'' effect \cite{Bliokh:09,Aiello:09}, the Orbital Angular Momentum (OAM) Hall effect \cite{Bliokh:06,Bekshaev:11}), or other ``spin-redirection'' phases \cite{Bliokh:15}.  Importantly, while the deflection of the beam centroid in these effects is typically on the order of a wavelength, the transverse displacement associated with our momentum exchange is many orders of magnitude larger and can be easily observed in experiment.  Indeed, the new transverse momentum component has been observed directly via its affect on the optical path (via Eq. \ref{eqn:velocity}), a prediction that is supported by our initial result in \cite{Nichols:22} and the results reported here.      
To our knowledge, such a model has not been previously put forward.  

\section{Experimental validation }\label{sec:experiment}

As noted in the previous sections, the influence of the polarization gradient momentum on a vector beam’s trajectory is directly observable through the proper choice of $\vec{\Omega}(\vec{x},0)$.  In \cite{Nichols:22}, it was shown that generating a vector beam with an initial ($z=0$) transverse polarization angle distribution of 
\begin{align}
\gamma(x,0)&=\frac{\pi}{2}\frac{(x-x_0)^2}{a^2}+\frac{\pi}{8}
\label{eqn:gamma}
\end{align}
will cause the beam to accelerate in the transverse direction along the chosen coordinate axis (in this case laboratory $x$). The parameters $a$ and $x_0$ determine the initial magnitude and direction of $\vec{\Omega}(x,0)$ and therefore the resulting curvature of the beam's path. Here, $x_0$ is measured relative to the centroid of $\rho(x,0)$.

\begin{figure}[tbh]
  \centerline{
   \begin{tabular}{c}
    \includegraphics[scale=0.75]{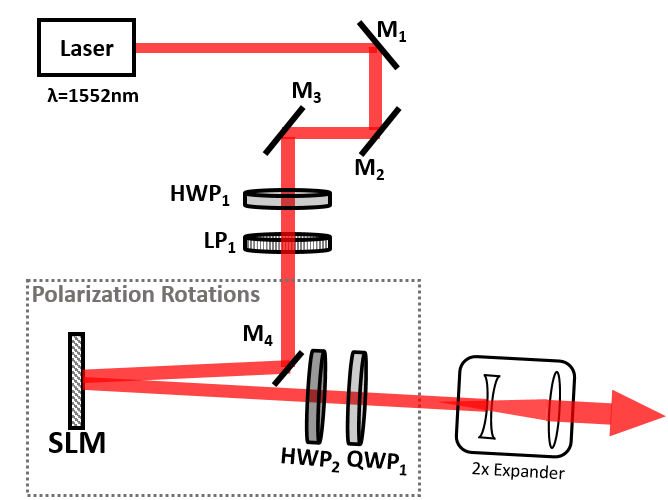}
  \end{tabular}
  }
  \caption{Experimental setup where a $\lambda$=1552nm beam with an initially spatially-homogeneous transverse linear polarization is transformed into a vector beam with a spatially-inhomogeneous polarization distribution using the series of polarization rotating elements enclosed by the gray rectangle (SLM $\rightarrow$ HWP$_2$ $\rightarrow$ QWP$_1$). Mirrors M$_1$-M$_3$ are used for beam alignment, whereas HWP$_1$ and LP$_1$ are used to ensure the correct input polarization.}
  \label{fig:setup}
\end{figure}

The optical setup used to create a vector beam with the desired $\gamma(x,0)$ is shown in Figure \ref{fig:setup}. First, after being reflected by a series of beam alignment mirrors (M$_1$-M$_3$), the output from a $\lambda$=1552nm laser is converted to a linearly polarized beam ( +45$^\circ$) by a half-wave plate (HWP$_1$) and linear polarizer (LP$_1$). After the LP$_1$, the beam is reflected by M$_4$ and is then incident on the spatial light modulator (SLM), which applies a {\it spatially-dependent} phase shift along the beam's $x$-axis. Consequently, the reflected beam's polarization state is no longer independent of its transverse spatial dimension(s), {\it that is}, it has become a vector beam. The vector beam is then transmitted through HWP$_2$, which applies another spatially-dependent polarization rotation (since the input beam's polarization is itself spatially-dependent). The beam then propagates through a quarter-wave plate (QWP$_1$), which applies the final polarization rotations. Finally, after QWP$_1$ the vector beam's width is expanded two-fold by a Galilean beam expander before propagating downrange. Note that in our implementation, the Gaussian beam width $w$ must be less than $a$ due to the width of the phase mask on the SLM and its maximum available phase retardation. The vector beam preparation technique, polarization rotations, and the resulting accrual of a spatially-varying PB phase is discussed in \cite{Nichols:22}.

\begin{figure}[tbh]
  \centerline{
   \begin{tabular}{c}
    \includegraphics[scale=0.32]{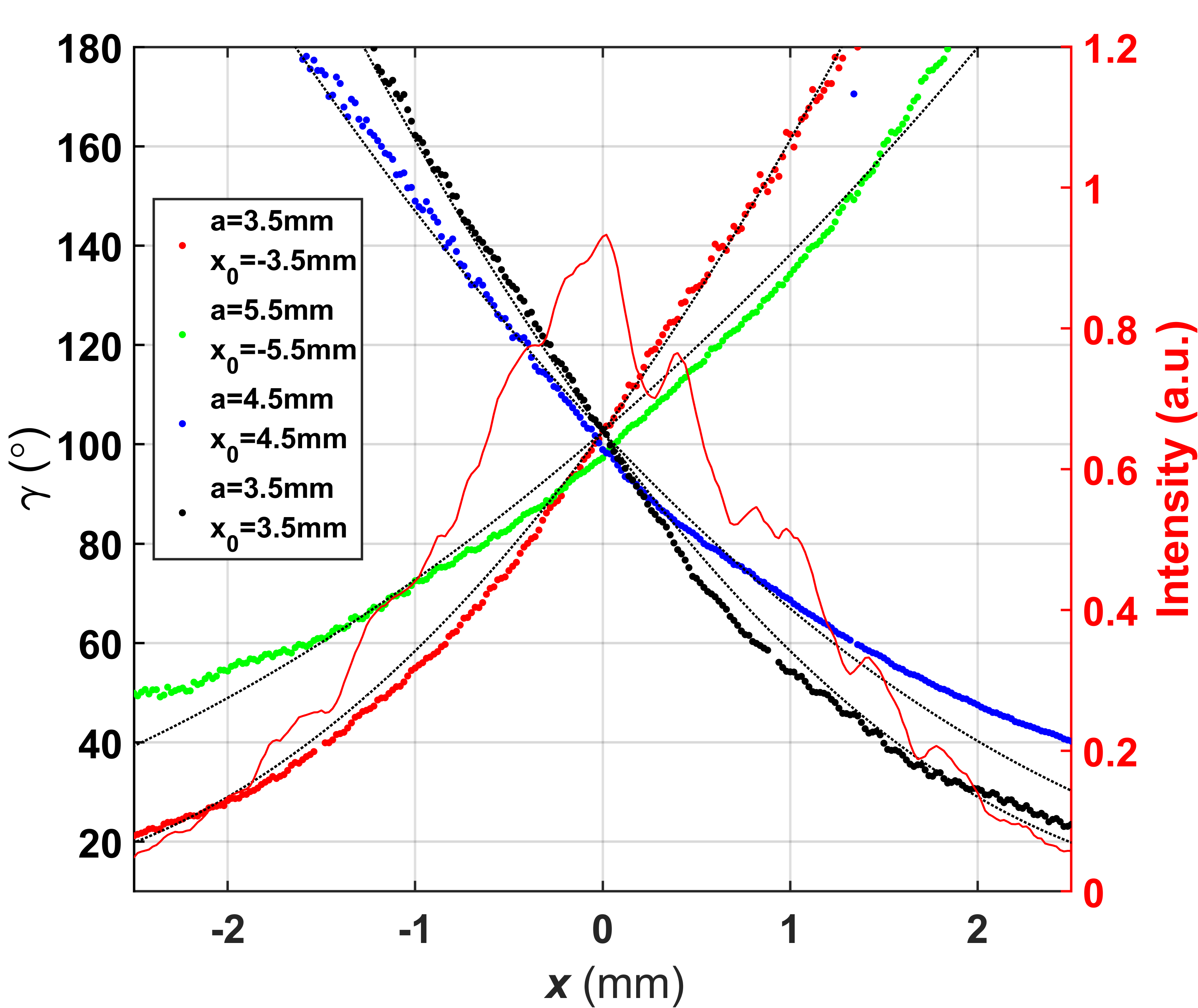}
  \end{tabular}
  }
  \caption{Left axis- measured transverse polarization angle distributions, $\gamma(x,0)$'s, of the vector beam with [$a$, $x_0$]'s of [3.5mm,-3.5mm] (red points), [5.5mm,-5.5mm] (green points), [4.5mm, 4.5mm] (blue points), and [3.5mm, 3.5mm] (black points). Also plotted are Eq. \ref{eqn:gamma} using the same values of [$a$, $x_0$] (black dotted lines). Right axis- an example measured intensity distribution, $\rho(x,0)$ (red line).}
  \label{fig:profiles}
\end{figure}

Figure \ref{fig:profiles} displays the measured transverse polarization angle distributions, $\gamma(x,0)$, at {\it{z}}=0 m (immediately after the output of the beam expander) for several values of $a$ and $x_0$. The different $\gamma(x,0)$'s were obtained by varying both parameters in Eq. \ref{eqn:gamma}, corresponding to variation of {\it{only}} the SLM's phase mask in the experimental setup. The $\gamma(x,0)$'s were measured using the Stokes polarimeter described in \cite{Nichols:22}. Also plotted in Figure \ref{fig:profiles} is a typical measured intensity profile, $\rho(x,0)$. The small fluctuations in $\rho(x,0)$ are simply measurement artifacts arising from Fabry-Perot reflections within the protective window of the InGaAs camera.

\begin{figure}[htb]
  \centerline{
   \begin{tabular}{c}
    \includegraphics[scale=0.325]{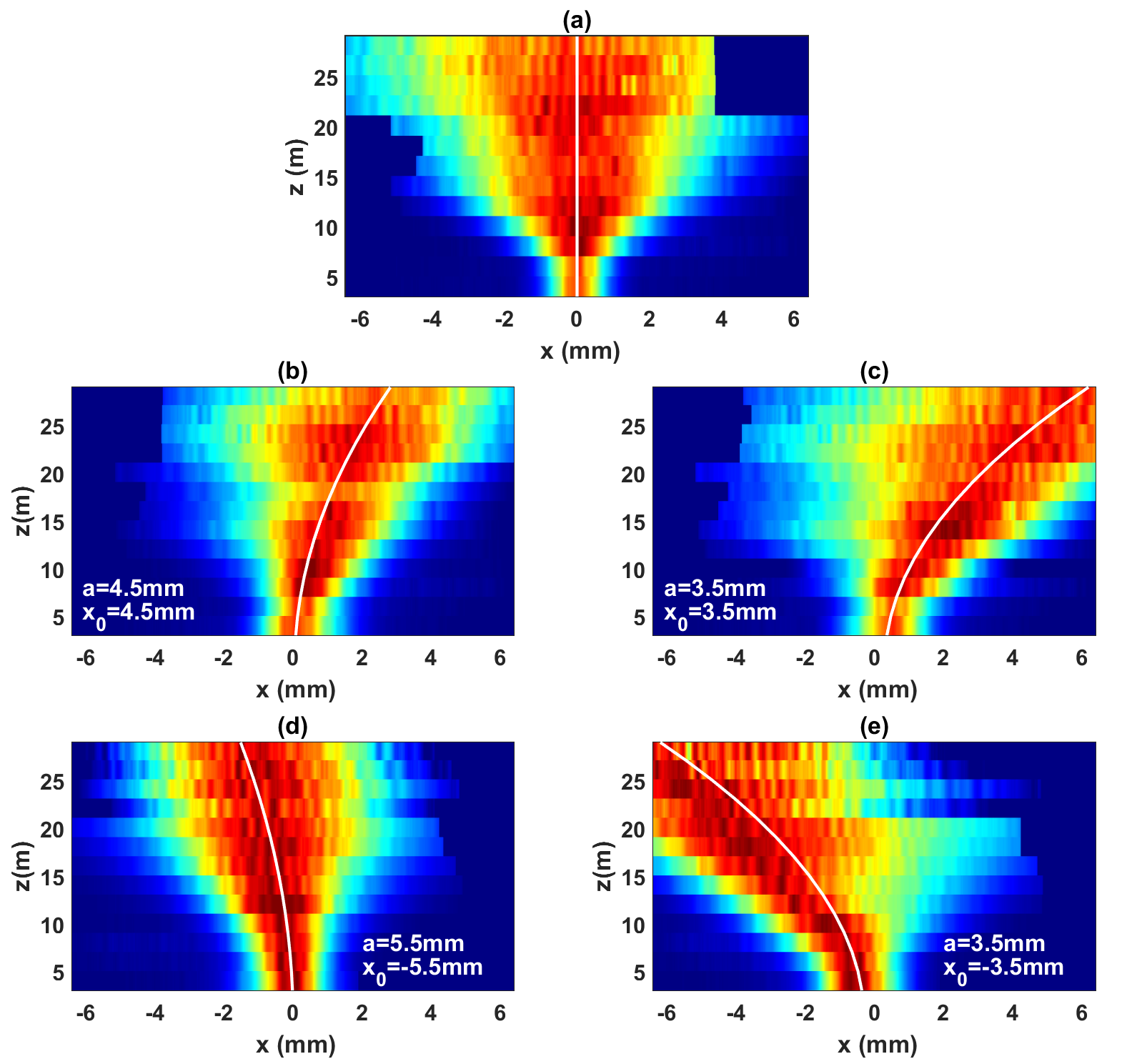}
  \end{tabular}
  }
  \caption{Normalized transverse intensity profiles, $\rho(x,z)$, of the beams vs. $z$. (a) $\rho(x,z)$ of a spatially-homogeneously, linearly polarized (scalar) beam. (b),(c),(d),(e): $\rho(x,z)$'s of the beams with [$a$, $x_0$]'s of [4.5mm,4.5mm], [3.5mm,3.5mm], [5.5mm,-5.5mm], and [3.5mm,-3.5mm], respectively. Also plotted are the centroids paths (white lines) predicted by Eq. (\ref{eqn:path})  using the same corresponding values of [$a$, $x_0$].}
  \label{fig:Iprofs}
\end{figure}

Figure \ref{fig:Iprofs} displays the normalized transverse intensity profiles of the vector beams measured at each different value of $z$ downrange ($\rho(x,z)$). Fig. \ref{fig:Iprofs} (a) shows the $\rho(x,z)$'s of a spatially-homogeneous, linearly polarized beam, which exhibits behavior that is well-described by standard Gaussian beam propagation. In contrast, Fig. \ref{fig:Iprofs} (b),(c),(d),and (e) display the $\rho(x,z)$'s of our vector beams with the initial $\gamma(x,0)$'s shown in Fig. \ref{fig:profiles}. The vector beam's clearly exhibit a nonlinear, or curved, path and the curvatures change as $a$ and $x_0$ are varied, with their centroids shifting by up to $\pm$6mm at $z$=30m. The paths of their centroids can be predicted by presuming that 1) diffraction acts as it does for a standard Gaussian beam, spreading light symmetrically about the centroid and that 2) diffractive effects do not appreciably reduce the polarization angle gradient via Eqn. (\ref{eqn:vorticity}). This latter assumption allows us to write $D\vec{\Omega}/Dz\approx 0$ and is akin to presuming an unlimited supply of stored momentum over the propagation range of interest.  With $\vec{\Omega}$ approximately constant in $z$, and a standard Gaussian diffraction model, Eqn. (\ref{eqn:velocity}) can be re-written to describe the evolution of beam center, $x_c\equiv 0$ at $z=0$.  Using Lagrangian coordinates (coordinates become functions of the independent variable $z$), $v_c(z)\equiv dx_c(z)/dz$ and
\begin{align}
    \frac{d^2 x_c(z)}{dz^2}&\approx -\left(\vec{\Omega}(x_c(0))\cdot\nabla_X\right)\vec{\Omega}(x_c(0))\nonumber \\
    &=\frac{\pi^2x_0}{2k_0^2a^4}
    \label{eqn:approx}
\end{align}
where we have used the polarization profile (\ref{eqn:gamma}) to obtain $\vec{\Omega}(x_c(0))$.  We can solve this expression with the initial conditions $dx_c(0)/dz=0$ (light is collimated) and $x_c(0)=0$.  Integrating Eqn. (\ref{eqn:approx}) twice in $z$ then gives for the beam center
\begin{align}
    x_c(z)\approx\frac{\pi^2x_0}{2k_0^2a^4}z^2.
    \label{eqn:path}
\end{align}
The paths predicted by Eq. \ref{eqn:path} are also plotted in Figure \ref{fig:Iprofs}, displaying good quantitative agreement with the measured profiles, with some deviations due to environmental effects. It is important to note that the transverse intensity profiles of the vectors beams remain well-described by a Gaussian distribution at each $z$ as they propagate downrange.

\begin{figure}[htb]
  \centerline{
   \begin{tabular}{ll}
    \includegraphics[scale=0.38]{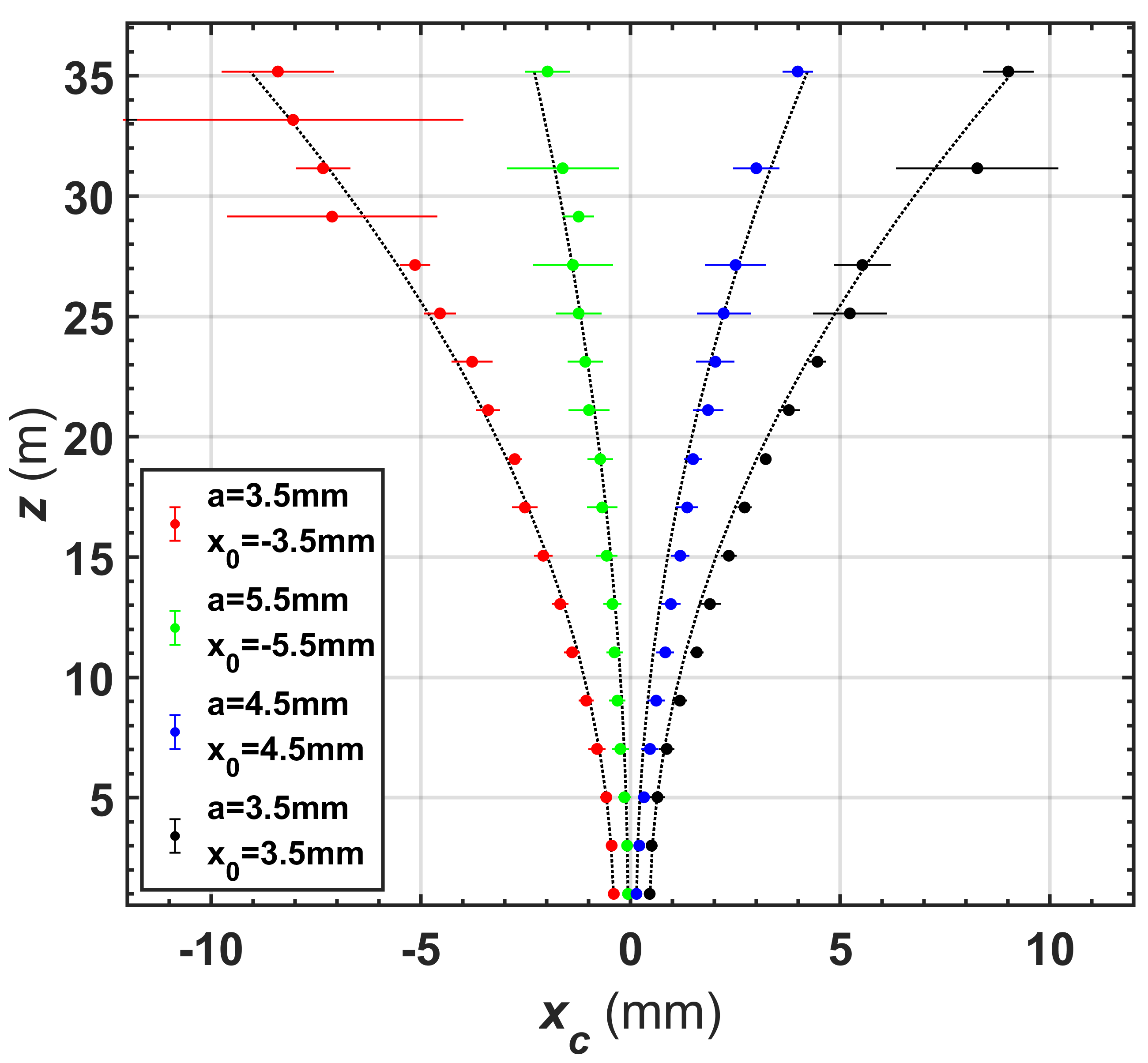}
  \end{tabular}
  }
  \caption{The measured transverse displacements of the vector beam centroids, $x_c$ vs. $z$, corresponding to the $\gamma(x,0)$'s of Fig. \ref{fig:profiles} with [$a$, $x_0$]'s of [3.5mm,-3.5mm] (red points), [5.5mm,-5.5mm] (green points), [4.5mm, 4.5mm] (blue points), and [3.5mm, 3.5mm] (black points). Also plotted are the approximate curves predicted by Eq. (\ref{eqn:path}) using the same corresponding values of [$a$, $x_0$].}
  \label{fig:displace}
\end{figure}

 Figure \ref{fig:displace} shows the measured transverse displacements of the beam centroids $x_c$ as a function of propagation distance out to $35$m along with their paths predicted by Eq. \ref{eqn:path}. The transverse displacements were measured using the procedure described in \cite{Nichols:22}, where the vector beam's displaced centroid position is measured relative to the centroid of the homogeneously polarized (scalar) reference beam at {\it{every}} value of $z$ downrange. In contrast to the single line images of $\rho(x,z)$  displayed in Fig. {\ref{fig:profiles}, here 10 images of {\it{both}} the reference beam and shifted vector beams are captured at each $z$ and their profiles averaged to reduce the influence of environmental fluctuations. With an initial Gaussian beam width of $w$=1.5mm and $a$'s between 3.5mm and 5.5mm, the predicted paths are in excellent agreement with the measured displacements over the range of $z$ values shown. The larger errors in $x_c$ beyond $z\approx$29m are most likely due to both an increase in beam pointing errors due to environmental fluctuations over longer path lengths and an increase in peak fit errors due to the beam's larger widths relative to the width of the camera. 
 \par It is important to understand properly the meaning of the data points in Fig.~\ref{fig:displace}. As already discussed, each data point simply represents the transverse location of the beam centroid as a function of downrange distance. It would be incorrect to conclude from the plot that angular momentum is not conserved as if the points somehow represented the motion of a particle with fixed mass traveling along a path with increasing curvature. The points represent only the positions of the intensity centroid. Upon propagation of the Gaussian beam both the intensity at the centroid and the overall polarization gradient decrease with increasing $z$ due to standard diffraction. As shown in Appendix (\ref{sec:conserve}), the quantity $\vec{P}=\rho\left(\vec{v}+\vec{\Omega}\right)$ captures the {\it local transverse momentum flux} and the total transverse momentum, given by the integral of $\vec{P}$ over the transverse plane, is independent of $z$, that is, it is conserved ( see Eq.~\ref{eqn:Reynolds}). Hence, as $\vec{\Omega}$ decreases with distance $z$, $\vec{v}$ must increase nonlinearly in order to conserve transverse momentum.

 \par We lastly note that although the approximation (\ref{eqn:path}) accurately predicts the results shown in Fig.~\ref{fig:displace}, our model suggests that diffraction will ultimately (at longer propagation distances) reduce the beam acceleration as discussed in Section (\ref{sec:momentum}). The coupling between diffraction and the polarization gradient and their mutual influence on the path of the beam are subjects of ongoing numerical and experimental investigations.

\section{Model extensions}

Although this paper is focused on free-space propagation, we can easily adapt the transport model to include propagation in a lossless, non-magnetic, isotropic, homogeneous medium characterized by relative permittivity $\epsilon_1(\vec{x},z)\ne 0$. This inclusion adds the term $k_0^2\epsilon_1(\vec{x},z)$ inside the brackets in (\ref{eqn:paraWave}) and, by extension, the term $\nabla_X\epsilon_1$ on the right hand side of Eq. (\ref{eqn:velocity}).  This term also results in the addition of the ``body force" 
\begin{align}
    \vec{s}(\vec{x},z)\equiv \rho(\vec{x},z)\nabla_X\epsilon_1(\vec{x},z)
\end{align}
to the right hand side of the conservation of momentum expression, Eq. (\ref{eqn:momentumEqn}).  This holds true whether the permittivity is assumed to be a deterministic function, or is rather capturing statistical variations in the refractive index of the medium \cite{Nichols:23}.

We also note that while we have considered coherent, fully polarized, monochromatic light, the model holds more generally for polychromatic, partially coherent, partially polarized light.  The only difference is that one has to re-define beam intensity, transverse phase gradient, and transverse polarization angle gradient as appropriate statistical quantities.  In the more general setting, these quantities were defined as averages over both time and spatial frequency in  \cite{Nichols:23}.  Appropriately, these quantities reduce to their coherent, monochromatic, fully polarized counterparts in those respective limits.  Thus, the model (\ref{eqn:model}) or (\ref{eqn:conserve}) represents a very general model for vector beam propagation under a wide variety of conditions. 

\section{Summary}

We have shown that paraxial beam propagation can in general be described by the model (\ref{eqn:model}).  The model is entirely consistent with the conservation equations found in many areas of continuum mechanics and can alternatively be written in the form (\ref{eqn:conserve}).  The model also clearly suggests a more general Poynting vector (\ref{eqn:Poynting}) which handles the spatially non-uniform polarization case, but reduces to the standard form for uniformly polarized light.  The acceleration of a beam in the transverse plane is viewed appropriately as a momentum exchange whereby an initial momentum, captured in the model by the polarization gradient, is converted to classical transverse momentum during propagation.  The physics of polarization gradient bending were alternatively described as a ``symmetry breaking'' in the Lagrangian density and as an observable manifestation of the vector potential found in other areas of physics.

An additional byproduct of the model is the interpretation of diffraction as a pressure.  The notion that intensity will always seek to move away from highly concentrated regions to flatter regions is consistent with both intuition and observation.   

We then validated the predictions of this model in experiment.  The vector beams were generated using sequential polarization rotations and thus are accompanied by a geometric, Pantcharatnam-Berry phase gradient.  Experimental measurements of the beam centroid as a function of propagation distance were then shown for different slopes and directions of the polarization gradient.  These results match closely the associated parabolic paths predicted by an approximate closed-form solution to the model.  The ultimate limits of the bending effect will be determined by both the strength of the polarization gradient, as this dictates the available momentum that can be exchanged for transverse motion of the centroid, and the initial width of the beam, which dictates the polarization gradient's rate of dissipation via diffraction.  More generally, the model provided here gives practitioners a fundamentally new design tool for tailoring the behavior of light beams.  New beam preparation approaches and testing longer propagation paths (numerically and experimentally) are the focus of ongoing investigations.

\section{Acknowledgments}
The authors would like to acknowledge support of the Office of Naval Research Codes 31 and 33 under grants N0001423WX01102, N0001422WX01660 

\iftrue

\appendix

\section{Derivation of the transport model for vector beams}\label{sec:derivation}
To derive the model (\ref{eqn:model}) and the ``conservation form'' (\ref{eqn:conserve}), we use the electric field model (\ref{eqn:efield}) with spatially varying polarization angle $\gamma(\vec{x},z)$ and substitute into Eq. (\ref{eqn:paraWave}). The result is the four mathematical statements:
\begin{subequations}
\begin{align}
k_0^{-1}\nabla_X\cdot\left[\rho(\vec{x},z)\nabla_X\gamma(\vec{x},z)\right]&=0
\label{eqn:sysA}
\end{align}
\begin{align}
\partial_z\rho(\vec{x},z)+k_0^{-1}\nabla_X\cdot\left[\rho(\vec{x},z)\nabla_X\phi(\vec{x},z))\right]&=0
\label{eqn:sysB}
\end{align}
\begin{align}
k_0\partial_z\phi(\vec{x},z)+\frac{1}{2}|\nabla_X\phi(\vec{x},z)|^2&+\frac{1}{2}|\nabla_X\gamma(\vec{x},z)|^2\nonumber \\
&=\frac{1}{2}\frac{\nabla_X^2\rho^{1/2}(\vec{x},z)}{\rho^{1/2}(\vec{x},z)}
\label{eqn:sysC}
\end{align}
\begin{align}
\frac{D\gamma(\vec{x},z)}{Dz}&=0,
\label{eqn:sysD}
\end{align}
\label{eqn:system}
\end{subequations}
where, recall, $D(\cdot)/Dz$ denotes the total derivative. Defining $\vec{v}(\vec{x},z)=k_0^{-1}\nabla_X\phi(\vec{x},x)$ and $\vec{\Omega}(\vec{x},z)=k_0^{-1}\nabla_X\gamma(\vec{x},z)$, (\ref{eqn:sysA}) can be added to (\ref{eqn:sysB}) to arrive at  (\ref{eqn:continuity},~\ref{eqn:continuityEqn}).  Taking the transverse gradient of both (\ref{eqn:sysC}) and (\ref{eqn:sysD}) yields, upon simplification, (\ref{eqn:velocity}) and (\ref{eqn:vorticity}).  In doing so, use is made of the identity $\nabla_X \left(\vec{A}\cdot\vec{B}\right)=(\vec{A}\cdot\nabla_X)\vec{B}+(\vec{B}\cdot\nabla_X)\vec{A}+\vec{A}\times(\nabla_X\times\vec{B})+\vec{B}\times(\nabla_X\times\vec{A})$. Since both $\vec{\Omega}(\vec{x},z)$ and $\vec{v}(\vec{x},z)$ are expressible as gradients of scalars, the cross-product terms vanish.  

To arrive at the ``conservation form'' of the model, first multiply (\ref{eqn:vorticity}) by $\rho(\vec{x},z)$ and add to the result the quantity
\begin{align}
&\left( \vec{v}(\vec{x},z)+\vec{\Omega}(\vec{x},z)\right)\nonumber \\
&\qquad\times\left(\vphantom{\vec{\Omega}}\partial_z\rho(\vec{x},z)+\nabla_X\cdot\left[\vphantom{\vec{\Omega}}\rho(\vec{x},z)\left(\vec{v}(\vec{x},z)+\vec{\Omega}(\vec{x},z)\right)\right]\right)
\end{align}
which, by  (\ref{eqn:continuity}), is identically zero.  With this addition, and use of the the vector identity $\nabla_X\cdot\left(\vec{B}\otimes\vec{A}\right)=\vec{A}\left(\nabla_X\cdot\vec{B}\right)+\left(\vec{B}\cdot\nabla_X\right)\vec{A}$, Eq. (\ref{eqn:vorticity}) becomes
\begin{align}   
    \partial_z&\left(\rho(\vec{x},z)\vec{\Omega}(\vec{x},z)\right)\nonumber \\
    &+\nabla_X\cdot\left[\rho(\vec{x},z)\left(\vec{v}(\vec{x},z)\otimes\vec{\Omega}(\vec{x},z)+\vec{\Omega}(\vec{x},z)\otimes \vec{v}(\vec{x},z)\right)\right]\nonumber \\
    =&-\vec{v}(\vec{x},z)\left(\vphantom{\vec{\Omega}}\partial_z\rho(\vec{x},z)+\nabla_X\cdot\left[\rho(\vec{x},z)\vec{v}(\vec{x},z)\right]\right)\nonumber \\
    &\quad -\vec{\Omega}(\vec{x},z)\nabla_X\cdot\left[\rho(\vec{x},z)\vec{\Omega}(\vec{x},z)\right].
    \label{eqn:step0}
\end{align}
Now multiplying Eq. (\ref{eqn:velocity}) by $\rho(\vec{x},z)$ and adding to Eq. (\ref{eqn:step0}) then gives the momentum conservation equation (\ref{eqn:momentumEqn}).  Thus, under the new, more general definition of the Poynting vector,  
\begin{align}
    \vec{P}(\vec{x},z)&=\rho(\vec{x},z)\left(\vec{v}(\vec{x},z)+\vec{\Omega}(\vec{x},z)\right),
\end{align}
the system (\ref{eqn:conserve}) express conservation of intensity and conservation of momentum for polarization gradient vector beams.

\section{Total Transverse Momentum Conservation \label{sec:conserve}}
Equations (\ref{eqn:continuityEqn}) and (\ref{eqn:momentumEqn}) are {\it local} expressions of conservation in the transverse plane in the sense that they depend on $\vec{x}$.  Using the Reynolds Transport Theorem (see e.g., \cite{Achenbach:73}, \cite{Marsden:83}) we can relate these expressions to global statements of conservation.  Specifically, Eq. (\ref{eqn:continuityEqn}) can be integrated in the transverse plane to yield
\begin{align}
    \frac{d}{dz}\int_{X_z}\rho(\vec{x},z)d\vec{x}&=0\label{eqn:conserveGlobal}.
\end{align}
In other words, the total intensity is conserved on propagation, although its {\it distribution} in the transverse plane may be altered under action of the field $\vec{v}(\vec{x},z)+\vec{\Omega}(\vec{x},z)$ via (\ref{eqn:continuity}). 

Likewise, Eq. (\ref{eqn:momentumEqn}) expresses the conservation of transverse linear momentum.  By integrating this expression in the transverse plane and again applying the Transport Theorem to the left hand side of (\ref{eqn:momentumEqn}) and the well known Divergence Theorem to the right hand side of Eq. (\ref{eqn:momentumEqn}), the expression can be written
 \begin{align}
 \frac{d}{dz}\int_{X_z} \vec{P}(\vec{x},z)d\vec{x}=-\int_{\partial X_z} {\bf P}\cdot\hat{n}~ds=0
 \label{eqn:Reynolds}
 \end{align}
where $ds=\sqrt{dx^2+dy^2}$ is the differential element along the closed curve $\partial_{X_z}$ which physically defines the boundary of the beam in the transverse plane as it propagates.  Along this boundary the intensity approaches zero by definition and therefore so too does ${\bf P}$ (by Eq. \ref{eqn:pressure}), hence diffraction does not change the total transverse linear momentum per unit length of propagation distance.  This was also the conclusion of \cite{Ghosh:83} in analyzing the analogous ``quantum pressure'' term found in hydrodynamic models of Schr\"{o}dingers equation \cite{Nore:93,Mocz:15}. 

Lastly, we note that we could have written these global, transverse conservation laws with $t$ as the independent variable by re-defining $\epsilon_0\rho(\vec{x},z)/c^2\sim [kg\cdot m^{-3}]$ (mass per unit volume) and $\epsilon_0\vec{P}(\vec{x},z)/c^2\sim [kg\cdot(m/s)\cdot m^{-3}]$ (transverse linear momentum density).  This alternative representation is realized by allowing $z\rightarrow c t$ and multiplying Eqs. (\ref{eqn:conserve}) by the free-space permittivity $\epsilon_0$.
\\*[0.1in]

\section{Polarization gradients and the Pancharatnam-Berry Phase \label{sec:vecPhase}} 


The electric field model (\ref{eqn:efield}) does not explicitly include the Pacharatnam-Berry (PB) phase that was acquired in creating the polarization gradient \cite{Nichols:22}.  However, at the aperture ($z=0$) the geometric PB phase and polarization angle are  directly related so that the subsequent evolution in the transverse gradient of the latter (Eq. \ref{eqn:vorticity}) can be viewed as predicting the transverse gradient of the former. 


To see this in more detail, we recall from our earlier work that the PB phase is generally a function of polarization angle, $F(\gamma(\vec{x}))$. For our particular experiment, this relationship is \cite{Nichols:22}
\begin{align} \phi_{PB}\left(\vec{x}\right)&=F(\vphantom{\vec{\Omega}}\gamma(\vec{x}))=-\frac{1}{2}\left(A\cos\left(\vphantom{\vec{\Omega}}2\gamma(\vec{x})\right)+B\right)
    \label{eqn:PBphase}
\end{align}
That is, during the process of creating the polarization angle gradient, the beam acquires a PB phase proportional to the cosine of twice the polarization angle.  Taking the transverse gradient of (\ref{eqn:PBphase}), rearranging, and multiplying by the intensity we can write
\begin{align}
  \rho(\vec{x},0)\vec{\Omega}(\vec{x},0)&=\frac{\rho(\vec{x},0)k_0^{-1}\nabla_XF\left(\vphantom{\vec{\Omega}}\gamma(\vec{x},0)\right)}{F'\left(\gamma(\vec{x},0)\right)}\nonumber \\
    &=\frac{\rho(\vec{x},0)k_0^{-1}\nabla_X\phi_{PB}\left(\vphantom{\vec{\Omega}}\vec{x}\right)}{A\sin\left(2\gamma(\vphantom{\vec{\Omega}}\vec{x},0)\right)}.
    \label{eqn:expression}
\end{align}
The polarization gradient is therefore proportional to a transverse PB phase gradient, a relationship also noted in \cite{Liu:15}, Eq. 4.  The factor in the denominator, $F'(\gamma(\vec{x},0))\equiv dF(\gamma(\vec{x}))/d\gamma(\vec{x})$, stems from applying the chain rule in differentiating (\ref{eqn:expression}) and depends in general on the particular sequence of transformations that lead to the polarization gradient.

The polarization gradient momentum could therefore also be described as a transverse momentum associated with a geometric phase gradient.  Put another way, our model is suggesting that the dynamic and geometric phases give rise to corresponding, distinct momenta. Once the the beam leaves the aperture, the transverse gradients $\vec{v}(\vec{x},z),~\vec{\Omega}(\vec{x},z)$ evolve according to (\ref{eqn:velocity}) and (\ref{eqn:vorticity}) respectively although they clearly remain coupled on propagation in such a way that the latter is eventually converted to the former causing the observed acceleration.

\fi 

\bibliography{main}  
\end{document}